\documentclass[aps,prl,twocolumn,showpacs,floatfix,reprint]{revtex4-1}
\usepackage[dvips]{graphicx}
\usepackage{color}
\usepackage{latexsym}
\usepackage{amsmath}
\usepackage{amsthm}
\usepackage{amsfonts}
\usepackage{amssymb}
\usepackage{mathrsfs}
\usepackage{natbib}
\usepackage{dsfont}

\begin{document}
\title{Ground-state cooling of a mechanical oscillator by interference in Andreev reflection}
\author{P. Stadler}
\author{W. Belzig}
\author{G. Rastelli}
\affiliation{Fachbereich Physik, Universit{\"a}t Konstanz, D-78457 Konstanz, Germany}
%

%
\begin{abstract}
%
%
We study the ground-state cooling of a mechanical oscillator linearly coupled to the charge of a 
quantum dot inserted between a normal metal and a superconducting contact.
Such a system can be realized e.g. by a suspended carbon nanotube quantum dot with  a  capacitive coupling
to a gate contact.
Focusing on the subgap transport regime, 
we analyze the inelastic Andreev reflections which drive the resonator to a non-equilibrium state.
For small coupling, we obtain that vibration-assisted reflections can occur through two distinct interference paths. 
The interference determines the ratio between the rates of absorption and emission of vibrational energy quanta. 
We show that ground-state cooling of the mechanical oscillator can be achieved for many of the oscillator's modes simultaneously or for single modes selectively, depending on the experimentally tunable coupling to the superconductor.
%
%
\end{abstract}
%

%
%
%
%
%
%
%

\date{\today}
\maketitle

%
%
%
%
%
%
{\sl Introduction.}- 
Nanoelectromechanical (NEMS) and optomechanical systems promise to manipulate mechanical motion in the 
quantum regime using, respectively, electrons \cite{Poot:2012fh,Greenberg:2012gi} or photons \cite{Aspelmeyer:2014ce}, 
for the realization of fundamental tests of quantum mechanics.
This goal requires the mechanical oscillator to be close to the quantum ground state, viz.
 $T \ll  \omega$, with $T$ the temperature and $\omega$ the mechanical frequency  $(\hbar=k_B=1)$.
Ground-state cooling, i.e. the average vibrational quanta $n \ll  1$, has been achieved in some nanomechanical devices, 
for instance in an oscillator of GHz frequency using standard dilution refrigeration \cite{OConnell:2010br}.
In another example, ground-state cooling was obtained in an opto-mechanical setup 
using the so-called side-band method \cite{Marquardt:2007dn,2008Sci...321.1172K} 
in which one mode of the resonator is coupled to a microwave electromagnetic cavity \cite{Teufel:2012jg}. 
Alternatively, several theoretical studies have analyzed  proposals for achieving cooling or ground-state cooling using electron transport 
\cite{Martin:2004kv,Lundin:2004cq,Rodrigues:2007kx,Brown:2007di,Xue:2007kx,Pistolesi:2009ks,Zippilli:2009gp,Galperin:2009fn,Sonne:2010jz,Santandrea:2011gh,Li:2011kn,Sonne:2011ii,Stadler:2014hra,Stadler:2015dga,Arrachea:2014cm}.
%
Most of them are closely related to the mechanism of the side-band cooling  \cite{Marquardt:2007dn,2008Sci...321.1172K}
and are based on an enhanced phonon absorption between two levels of energy difference $\Delta E$. 
As consequence, cooling is expected when the resonant condition $\omega = \Delta E$  is satisfied. 

So far cooling by electron transport has been experimentally reported  
in a resonator coupled to a superconducting single electron transistor \cite{Naik:2006gf,Blencowe:2005ih,Clerk:2005ei}.
%
%
%
%
Furthermore, suspended carbon nanotube quantum dots (CNTQDs) have proved to be ideal candidates for quantum NEMS 
\cite{Huttel:2009jd,Moser:2014je,Benyamini:2014eb}   but 
the typical frequencies of the relevant modes ($f \alt 100 $MHz) correspond to demanding 
cooling temperatures for electronic circuits \cite{Huttel:2009jd}. 

\begin{figure}[tbp]
	\includegraphics[scale=0.65,angle=0.]{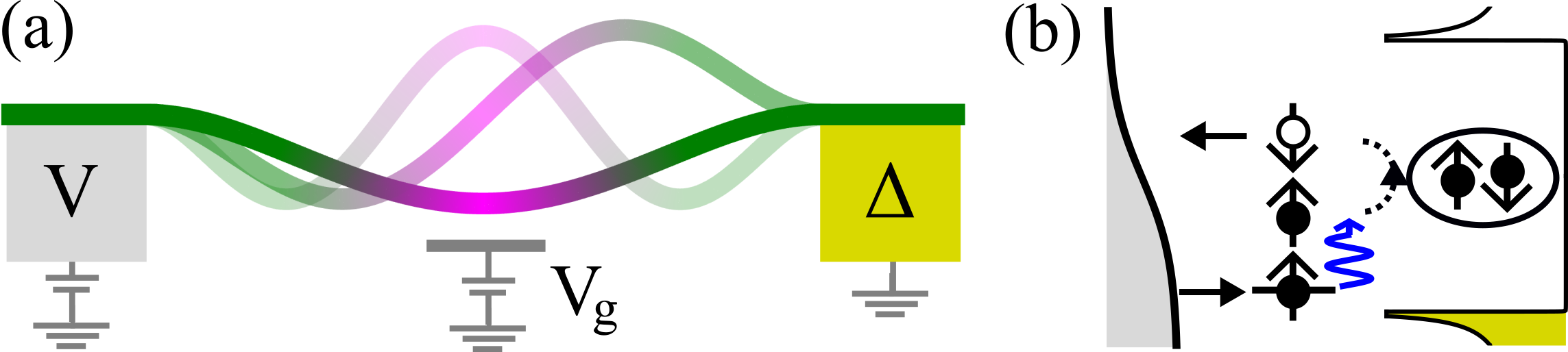}
	\caption{	(a) Suspended CNTQD between a normal lead and a superconductor with a coupling 
			between the dot's charge and the flexural mechanical modes.
			(b) Example of inelastic AR: An incoming electron absorbs one  phonon 	
			from the resonator  (blue arrow) before to be reflected as hole.	}
	\label{fig:schema_system_idea}
\end{figure}

In this Letter, we analyze ground-state cooling of a CNTQD suspended 
between a normal metal and a superconductor, Fig.~\ref{fig:schema_system_idea}(a).
In the subgap transport regime  $|eV| \ll \Delta$ - with $V$ the voltage and $\Delta$ the superconducting gap - 
we found ground-state cooling due to inelastic Andreev reflections (ARs), Fig.~\ref{fig:schema_system_idea}(b),
without the requirement of any resonant condition.

%
%

\begin{figure}[htbp]			
		\includegraphics[scale=0.38,angle=0.]{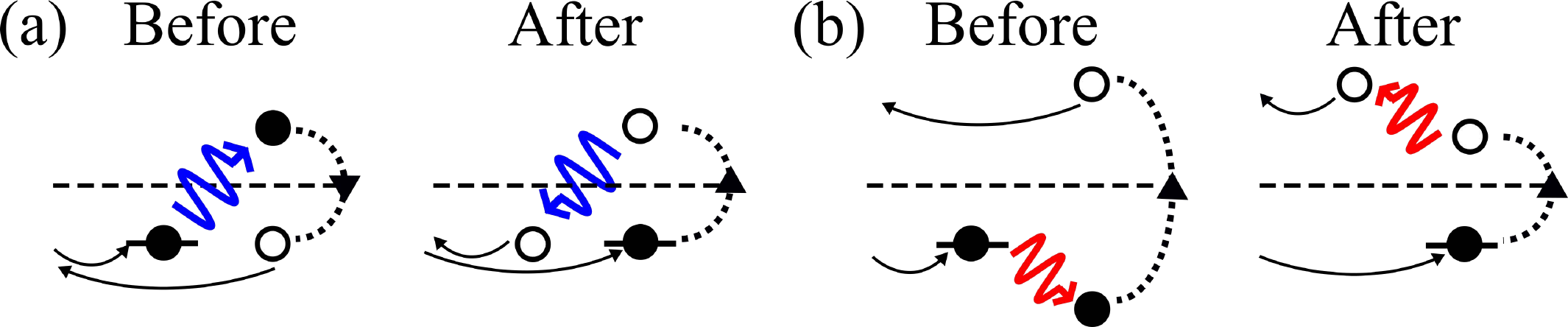}  
	\caption{Inelastic ARs for an incoming electron from the normal lead with the dot's energy 
	 below  the Fermi level of the  superconductor  $(\mu=0)$ (dashed line).  
		(a) The two possible paths in which the electron absorbs a phonon 
		(blue arrow)  {\sl before}  or  {\sl after} the AR (dotted line).
		(b) The two possible paths for the electron emitting a phonon (red arrow). }  
	\label{fig:diffsplitting}
\end{figure}

In an AR, an incident electron from the normal contact 
forms a Cooper pair in the superconductor  with the reflection of a hole. 
%
Due to the interaction with the mechanical oscillator, 
ARs can be vibration-assisted with the absorption or emission of a vibrational energy quantum or phonon, see 
Fig.~\ref{fig:schema_system_idea}(b). 
For weak coupling,  inelastic ARs involve only one phonon at a time and 
have two possible paths associated with the energy exchange with the resonator 
before or after an AR, see  Fig.~\ref{fig:diffsplitting}.
These two paths can interfere. 
Hence, by varying the dot's energy level,  we can achieve destructive interference between the paths 
associated to the phonon emission, Fig.~\ref{fig:diffsplitting}(b),  such that the resonator is cooled since
the ARs with phonon absorption dominate, Fig.~\ref{fig:diffsplitting}(a). 
This destructive interference can occur 
in a wide frequency range, which allows to achieve simultaneous
ground-state cooling  of multiple mechanical modes.

%
%
%

%
%
%
%
%
%
%
%
%
%
{\sl  Model.}- 
%
We consider the Hamiltonian  $\hat{H}=\hat{H}_n+\hat{H}_{t}+\hat{H}_{dS}+\hat{H}_m$.  
The part of the normal lead and its tunnel coupling with the quantum dot reads 
$\hat{H}_n + \hat{H}_t = \sum_{k\sigma} [ (\varepsilon_k  - eV)  \hat{c}_{k\sigma}^\dagger \hat{c}_{k\sigma}^{\phantom{\dagger}} + 
(t_n \hat{d}^{\dagger}_{\sigma}  \hat{c}_{k\sigma} +  \textrm{H.c.} ) ]$
with $\hat{c}_{k,\sigma}$ and $\hat{d}_\sigma$ the annihilation operators for the 
electronic states $k$ and spin $\sigma$ in the normal lead and in the dot.  
The tunneling rate is $\Gamma_n=\pi \rho_n {|t_n|}^2$ 
with $\rho_n$ the density of states. 
Elastic as well as inelastic tunneling of quasiparticles above the gap can be neglected in the deep subgap regime 
$|eV| \ll \Delta$ as their contribution is exponentially small in $\Delta/T$ for 
$T\ll \Delta$ (see Sup.~Mat.~\cite{SupMat} and Ref.\cite{Cuevas:1996tm}).
%
Therefore, we study the following effective Hamiltonian for the dot $\hat{H}_{dS} =  \sum_{\sigma} 
\varepsilon_{0}^{\phantom{\dagger}}  \hat{d}^{\dagger}_{\sigma} \hat{d}_{\sigma}^{\phantom{}} 
- \Gamma_s ( \hat{d}^{\dagger}_{\uparrow}  \hat{d}^{\dagger}_{\downarrow} + \hat{d}_{\downarrow}^{\phantom{}} \hat{d}_{\uparrow}^{\phantom{}} )$
with $\varepsilon_{0}$ the dot's energy for two spin-degenerate levels and 
 $\Gamma_s$ the coupling strength for the intra-dot  pairing due to the proximity with the superconductor.
The eigenstates of  $\hat{H}_{dS}$
correspond to Andreev  states  
formed by coherent superposition of electrons and holes with energies $\pm E_A= \pm \sqrt{\varepsilon_0^2+\Gamma_s^2}$
\cite{no-broadening_Gamma_S}. 

In a suspended CNTQD, the electrostatic force between the nanotube and the gate leads to a 
capacitive coupling between the flexural modes and the dot's charge  \cite{Huttel:2009jd}.
%
%
Expanding the electrostatic energy in 
terms of the tube's transversal displacement and the average dot's charge 
leads to an Holstein  interaction 
$\hat{H}_{m} = \sum_k [ \omega^{\phantom{g}}_k  \hat{b}_k^{\dagger} \hat{b}^{\phantom{g}}_k + 
\lambda^{\phantom{g}}_k  (\hat{b}_k^{\phantom{g}} + \hat{b}^{\dagger}_k) \hat{n}_d ]$ 
with $\hat{b}_k$  the bosonic annihilation operators for the flexural modes $(k=1,2,\dots)$ 
of frequency $\omega_k$ and  $n_d$ is the fluctuating part of the charge \cite{Braig:2003cb,Rastelli:2012dh}.  
Assuming weak coupling, one can neglect the effects of the resonator on the electron system in a first approximation.
Then we can analyze the electromechanical damping rate $\gamma$ and
the non-equilibrium phonon occupation number $n$ due to the charge tunneling,  
separately for each mechanical mode.
In the next two sections, we discuss the single-mode case.

%
%
%
%
%
%
%
%
%
%
{\sl  Damping for a single mode.}-
We found that the electromechanical damping is determined by inelastic ARs and normal reflections (NRs).
Detailed calculations are reported in~\cite{SupMat}.
For the damping rate we obtain the result $\gamma =  \gamma_{\text{AR}} + \gamma_{\text{NR}}$ 
in which, for instance, the damping associated to the ARs reads  
%
%
%
%
%
%
%
%
%
%
%
\begin{equation}
\label{eq:gamma_AR}
\gamma_{\text{AR}} =  \gamma_{eh}^{+}  +  \gamma_{he}^{+}  - \gamma_{eh}^{-}  -  \gamma_{he}^{-}  \, . 
\end{equation}
The individual rates $ \gamma_{eh}^{s}$ and $\gamma_{he}^{s}$  in Eq.~(\ref{eq:gamma_AR}) correspond to inelastic AR with the absorption 
$\!s\!=\!+$ or emission $\!s\!=\!\!-$ of one phonon for an incoming electron from the normal lead  ($eh$) or an incoming hole ($he$).
%
As example, the rates for an electron reflected as a hole ($eh$) take the form
%
%
%
%
%
%
%
%
%
%
%
\begin{equation}
\label{eq:rates_reflections}
\! \gamma_{eh}^{\pm} \!=\!   \frac{  \lambda^2 \Gamma_n^2}{2} \!\!
\int \!\! \frac{d\varepsilon}{2\pi} \, 
f_e(\varepsilon)  [1-f_{h}(\varepsilon\pm\omega)] 
\, 
\left| A_{\pm}(\varepsilon) \!+\! B_{\pm}(\varepsilon)   \right|^2  , 
\end{equation}
with $\lambda$ the charge-vibration coupling constant and the Fermi functions 
$f_{e}(\varepsilon) = {\left[ 1 + \exp\left( (\varepsilon-\, eV)/T \right) \right]}^{-1} $ for the electrons 
and  $f_{h}(\varepsilon)= 1 - f_e(-\varepsilon)$ for the holes.
%
%
%
Hereafter, to be definite, we consider $eV>0$ and the high-voltage limit, namely $eV  \gg  T, \omega, E_A$. 
In this case the rates of the reflections for incoming holes are negligible compared to the ones  
associated to electrons  (vice versa for $eV<0$) and we approximate 
$\gamma_{\textrm{AR}} \simeq  \gamma_{eh}^{+}   - \gamma_{eh}^{-} $.
Moreover, we can approximate $f_{e}\simeq 1$ and $f_{h}\simeq 0$  in Eq.~\eqref{eq:rates_reflections}.
Hence, the behavior of the rates $\gamma_{eh}^{\pm}$ are ruled solely by the  last term inside the integral Eq.~\eqref{eq:rates_reflections} 
that represents the transmission for inelastic ARs of an incoming electron.
The transmission is given by the coherent sum of two amplitudes that are associated to the two possible paths  in 
which the phonon is emitted or absorbed before $(B_{\pm})$  or after $(A_{\pm})$ a single AR, Fig.~\ref{fig:diffsplitting} (see also~\cite{SupMat}).
%

%
%
%
%
%
%
%
%
%
%
{\sl Phonon occupation due to inelastic ARs.}
We first analyze the contribution of inelastic ARs to the phonon occupation assuming that NRs are negligible.
In this case we find the  result
%
%
%
%
%
%
%
%
\begin{equation}
\label{eq:ratio_gammas}
n=n_{\textrm{AR}} \equiv  \frac{1}{\kappa  - 1} \, , \,\,\, 
\kappa = \frac{\gamma^{+}_{eh}}{\gamma^{-}_{eh}} \simeq  
\, \frac{(\varepsilon_0 - \omega/2)^2 + \Gamma_n^2}{(\varepsilon_0+\omega/2)^2 + \Gamma_n^2}  \, ,
\end{equation}
valid for $\kappa > 1$. 
Thus, $n_{\textrm{AR}}$ is the outcome of the competition between emission and absorption processes for inelastic ARs:
for $\kappa \gg 1$,  the resonator can be cooled to the ground state with $n_{\textrm{AR}} \ll 1$, 
whereas the phonon occupation is increased for $\kappa \agt 1$ such that $n_{\textrm{AR}} \gg 1$. 
Eventually the resonator is unstable for $\kappa < 1$  \cite{Stadler:2015dga}.
The different regimes can be reached only by tuning the dot's energy level $\varepsilon_0$ (i.e. 
the gate voltage):  $n_{\textrm{AR}} > 1$ and the instability always occur  for $\varepsilon_0 > 0$ whereas
$n_{\textrm{AR}}<1$ is achieved as long as  $\varepsilon_0 < 0$.
In particular, the lowest phonon occupation is given by $n_{\text{opt}} = {(\Gamma_n/\omega})^2 $ 
corresponding to ground-state cooling for $\Gamma_n \ll \omega$ \cite{Benyamini:2014eb}.

\begin{figure}[b]
	\begin{flushleft}
		\begin{minipage}{0.45\columnwidth}
			\includegraphics[scale=0.4,angle=0.]{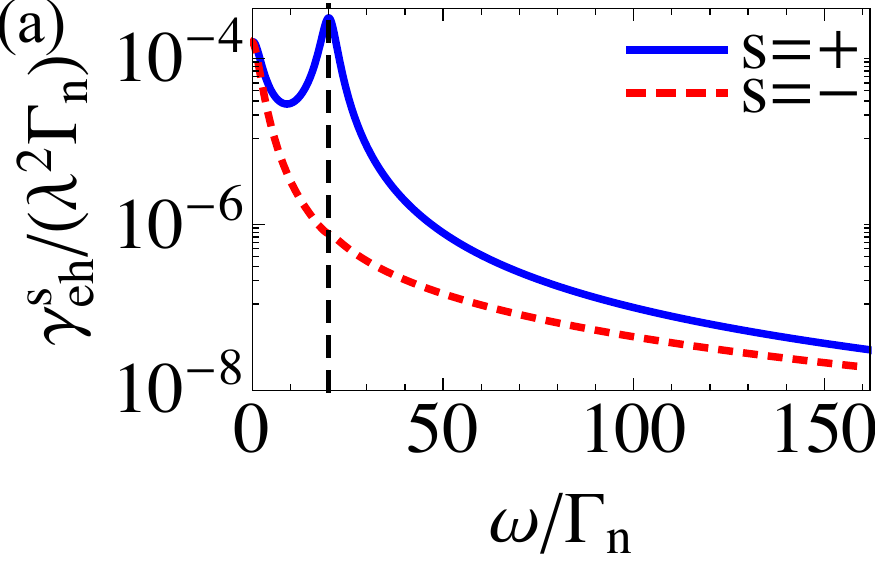}  
		\end{minipage} 
		\begin{minipage}{0.45\columnwidth}
			\includegraphics[scale=0.4,angle=0.]{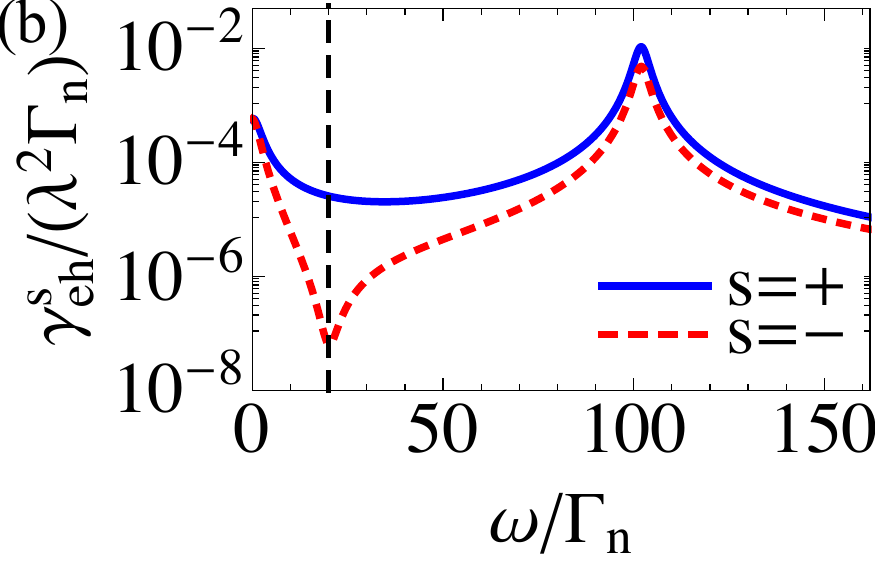}	
		\end{minipage}
	\end{flushleft}		
	\caption{The rates  
	         $\gamma^{\pm}_{eh}$ for inelastic ARs for $\varepsilon_0=-10\Gamma_n$ (s=+ absorption, s=- emission).
	         The vertical dashed line is $\omega=2\vert \varepsilon_0 \vert$. 
		(a) Weak coupling regime between the dot and the superconductor with $\Gamma_s = 0.1 |\varepsilon_0|$ 
		(b) Strong coupling regime with $\Gamma_s =5 |\varepsilon_0|$. }    
	\label{fig:noise}
\end{figure}

It is interesting to analyze the behavior of the individual rates  as a function of $\omega$, Fig.~\ref{fig:noise}.
Focusing on the regime $\varepsilon_0<0$,  
%
the condition $\kappa \gg 1$  for the ground-state cooling   
occurs either in the limit $\Gamma_s \ll |\varepsilon_0|$ when $\gamma^{+}_{eh}$  has a peak,  Fig.~\ref{fig:noise}(a), 
or in the limit $\Gamma_s \gg |\varepsilon_0|$ when $\gamma^{-}_{eh}$  has a broadened depletion around a dip,  Fig.~\ref{fig:noise}(b).

%
%
%
%
%

The peak in  Fig.~\ref{fig:noise}(a) results from a resonance:
the reflected hole is injected at the same energy as the incoming electron $\varepsilon_0=-\omega/2$ in the case of phonon absorption,  Fig.~\ref{fig:diffsplitting}(a).
This energy  alignment holds independently whether the absorption occurs after or before an AR 
and enhances the phonon absorption amplitudes. 
Such an alignment  does not occur for ARs with phonon emission, Fig.~\ref{fig:diffsplitting}(b). 

In contrast, the dip of the emission rate in Fig.~\ref{fig:noise}(b) occurs because the two paths with phonon emission 
in Fig.~\ref{fig:diffsplitting}(b) interfere destructively,  $|A_{-}+B_{-}| \ll |A_-|,|B_-| $ as mentioned  earlier,    
and the resonator is cooled due to the absorption process, namely $|A_+ + B_+| \gg |A_-+ B_-|$. 
The latter condition occurs even when the moduli of all the amplitudes 
are of the same order of magnitude. 
Eventually, increasing the frequency,  the two rates  become comparable $\gamma_{eh}^{+} \sim  \gamma_{eh}^{-}$ and 
both rates show a peak at $\omega \simeq 2 \Gamma_S$, Fig.~\ref{fig:noise}(b),  corresponding to the  energy separation between 
the two Andreev levels  $(E_A\simeq\Gamma_s)$. 
However, for $|\varepsilon_0| \, \alt \omega \, \alt \Gamma_S$, we notice $\gamma_{eh}^{-}$ 
is still two orders of magnitudes smaller than $\gamma_{eh}^{+}$, Fig.~\ref{fig:noise}(b).
This suppression of the emission rate for inelastic AR in a wide frequency range 
allows to cool many mechanical modes of different frequency.
The extension of this suppression  sets approximately the cooling spectral band.

Such an interference mechanism of cooling is different from the method based on the quasi-particles transport   
involving inelastic tunnelling with electronic states  above the gap \cite{Hekking:2008hx,Koppinen:2009bn,Feshchenko:2015kn}. 
Inelastic quasi-particles and Andreev transport for heating has been discussed in \cite{Wang:2013kc} although 
subgap ground-state cooling was not studied. 
Cooling by interference was also discussed in ~\cite{Elste:2009hf}  but by using a dissipative optomechanical coupling.

\begin{figure}[tbp]
	\includegraphics[scale=0.3,angle=0.]{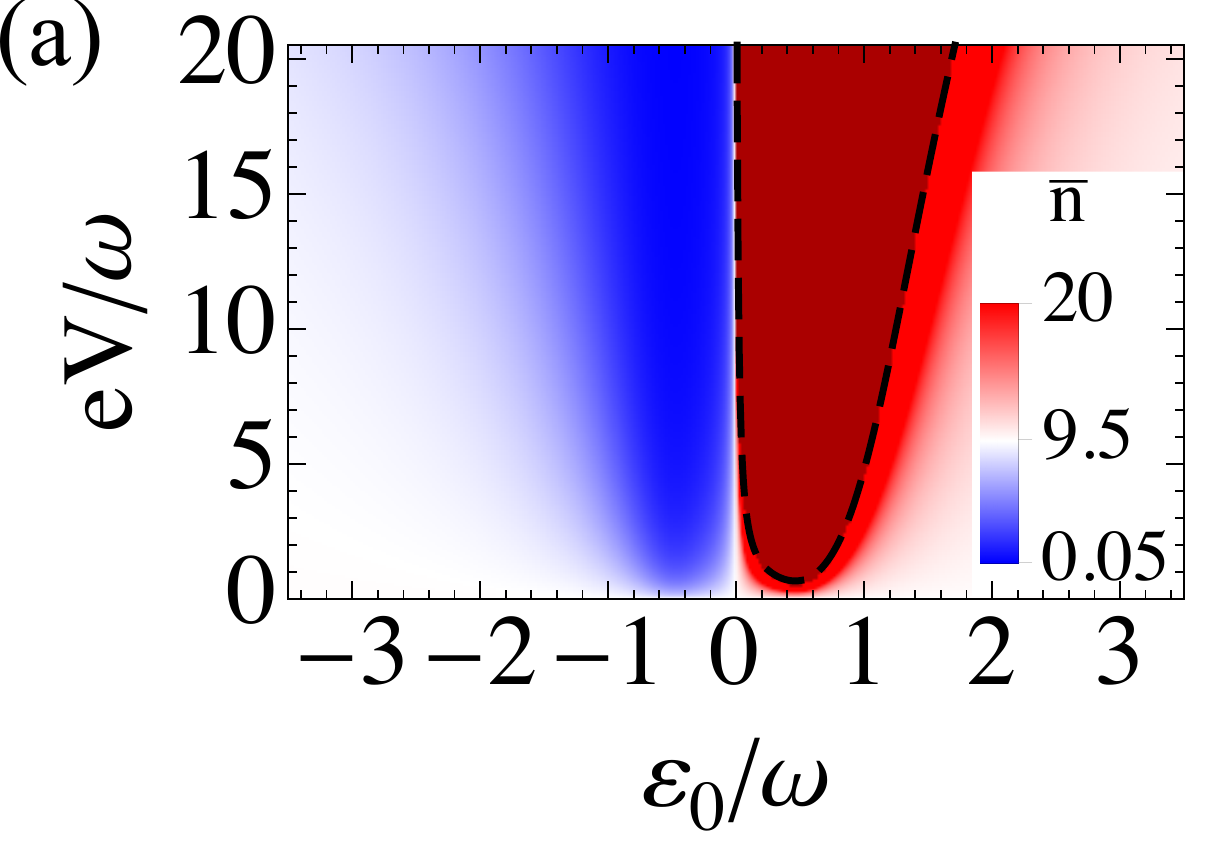}
	\includegraphics[scale=0.42,angle=0.]{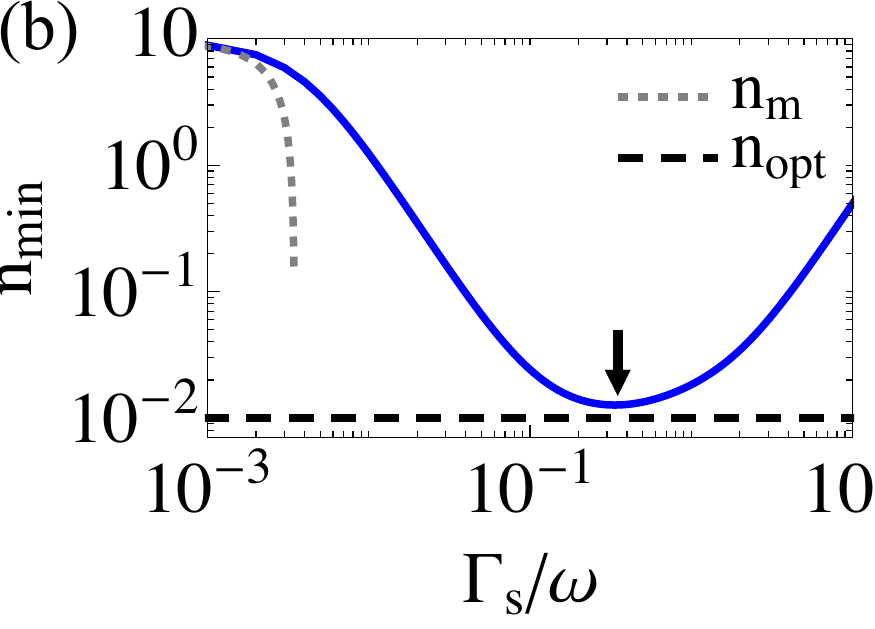}
	\caption{ (a) Phonon occupation as a function of $V$ and $\varepsilon_0$ for $\Gamma_s=0.35\omega$. 
		White color corresponds to $n_B(\omega)$. The dark red region limited by the dashed line corresponds to the instability region 
		$\gamma_{\textrm{AR}} + \gamma_{\textrm{NR}}  +\gamma_0<0$.  
		(b) Minimal phonon occupation as a function of $\Gamma_s$ for $\varepsilon_0=-\omega/2$.
		The lowest possible occupation  is $n_{\textrm{opt}}={(\Gamma_n/\omega)}^2$ (dashed black line). 
		The dotted line (gray) is the analytic approximation 
		$n_{m} \approx n_B(\omega) [ 1 -  8\pi \Gamma^2_s/(\Gamma_n \gamma_{\textrm{NR}})]$
		for   $\Gamma_s \ll \omega$, see also  \cite{SupMat}.
		The arrow points the optimal point at $\Gamma_s=0.35\omega$. 
		Parameters: $\Gamma_n = 0.1\omega$, $T= 10 \omega $, $\gamma_0/\omega = 10^{-6}$ and $\lambda=0.1\omega$  \cite{Rastelli:2012dh}.}    
	%
	%
	\label{fig:diffsplitting2}
\end{figure}

%
%
%
%
%
%
%
%
%
%
{\sl General results and effects of the normal reflections.}-
Formally, the electromechanical damping $\gamma$ and the steady non-equilibrium phonon occupation $n$ 
are determined by the spectrum of the non-symmetrized noise of the dot's charge   occupation  
$ S(\varepsilon) = \int dt \, e^{i \varepsilon t} {\left<  \hat{n}_d(t) \hat{n}_d \right>}$  where the 
quantum statistical average is taken over the electron system \cite{Clerk:2004cz}.  
%
%
Explicit relations between $\gamma$, $n$ and the noise $ S(\varepsilon)$  are given in ~\cite{SupMat}. 
For the phonon occupation, we obtain  
%
%
%
%
%
%
%
%
%
\begin{equation}
\label{eq:main_result}
\bar{n} = \frac{\gamma_{\textrm{AR}}\,\,n_{\textrm{AR}}+ \left(\gamma_{\textrm{NR}}+\gamma_0\right) n_B(\omega) }{\gamma_{\textrm{AR}}+\gamma_{\textrm{NR}}+\gamma_0}  \, ,
\end{equation}
with the Bose function  $n_B(\omega)= {\left[\exp\left( \omega /T \right)-1 \right]}^{-1}$ and an intrinsic damping $\gamma_0=\omega/Q$ , with the quality factor 
$Q\!\!\sim\!\!10^{6}$  \cite{Huttel:2009jd,Moser:2014je}. 
%
The NRs that involve only the normal lead at the bath temperature 
can drive the oscillator only towards thermal equilibrium.
The general expression for $n_{\mathrm{AR}}$ reads 
%
%
%
%
%
%
%
%
%
\begin{align}
\!\!n_{\textrm{AR}} \!= \!\!\sum_{s=\pm} \! \left[ \gamma^{s}_{eh} n_B\left( \omega \!+\!  s \, 2eV \right) 
\!+\! \gamma^{s}_{he} n_B\left( \omega \!-\!  s \, 2eV \right) \right]\! /\! \gamma_{\textrm{AR}} \, ,
\end{align}
An example of the result  for $\bar{n}$ is shown in Fig.~\ref{fig:diffsplitting2}(a) for some realistic parameters.
Restoring the NRs in the phonon occupation increases the  minimum occupation attainable by ARs.
However, we obtain $n_{min} \simeq 0.05$ in the region $\varepsilon_0<0$  of Fig.~\ref{fig:diffsplitting2}(a), i.e. ground state cooling is  still feasible. 
In the region of the dot's level $\varepsilon_0>0$ the situation in Fig.~\ref{fig:diffsplitting2}(a) is inverted: 
the emission rates of the inelastic ARs dominates over the absorption ones leading to increase of the phonon occupation (in the    
region of stability $\gamma_{\textrm{AR}} + \gamma_{\textrm{NR}}  +\gamma_0>0$),  and eventually to a mechanical instability.

%
%
%
%
%

When we take into account NRs the minimal occupation becomes a function of $\Gamma_s$ and there is 
an optimal value for the coupling with the superconducting lead that maximizes the cooling, as 
shown in Fig.~\ref{fig:diffsplitting2}(b). 
Setting $\varepsilon_0 = -\omega/2$ and $eV \gg (\omega ,T)$, 
in the limit  of small intrinsic damping $\gamma_{0} \ll \gamma_{\textrm{NR}}$ 
and for strong suppression of the phonon emission rate $\gamma^{-}_{eh} \ll \gamma^{+}_{eh} $ ($n_{\text{opt}} \ll  1$),  
we have 
$n_{min} \approx 
 \left(
 \gamma^{+}_{eh} n_{\textrm{opt}} + \gamma_{\textrm{NR}} n_B(\omega) 
 \right)
 /( \gamma^{+}_{eh} +\gamma_{\textrm{NR}} )$. 
For  $\Gamma_s  \rightarrow 0 $,  we have $\gamma^{+}_{eh}\rightarrow 0 $, 
NRs dominate over ARs and the oscillator is close to the thermal equilibrium. 
Increasing $\Gamma_s$, the resonator starts to be cooled due to the ARs and the phonon occupation approaches the optimal value $n_{\textrm{opt}}$. 
As the AR rate $\gamma_{eh}^+$ vanishes at large $\Gamma_s$
(see Fig.~\ref{fig:noise}),  $n_{\textrm{min}}$ shows a non monotonic behavior.

%
%
%
%
%
%
%
%
%
{\sl Results for several mechanical modes.}-
%
%
In this section we illustrate the possibility of cooling several non-degenerate mechanical modes  
owing to the interference between the inelastic ARs paths with phonon emission.  

\begin{figure}[tbp]
	\includegraphics[scale=0.25,angle=0.]{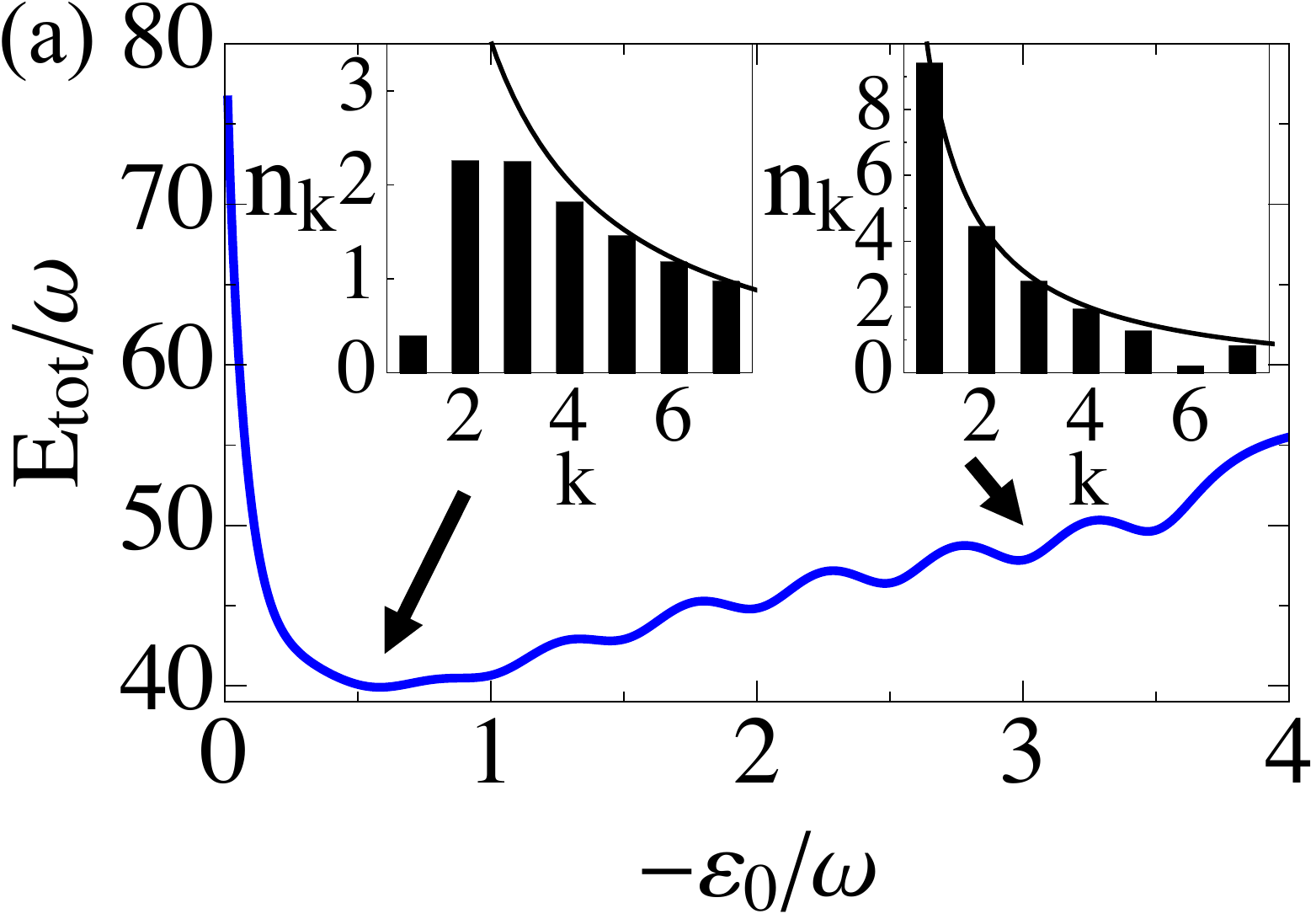} 
	\includegraphics[scale=0.25,angle=0.]{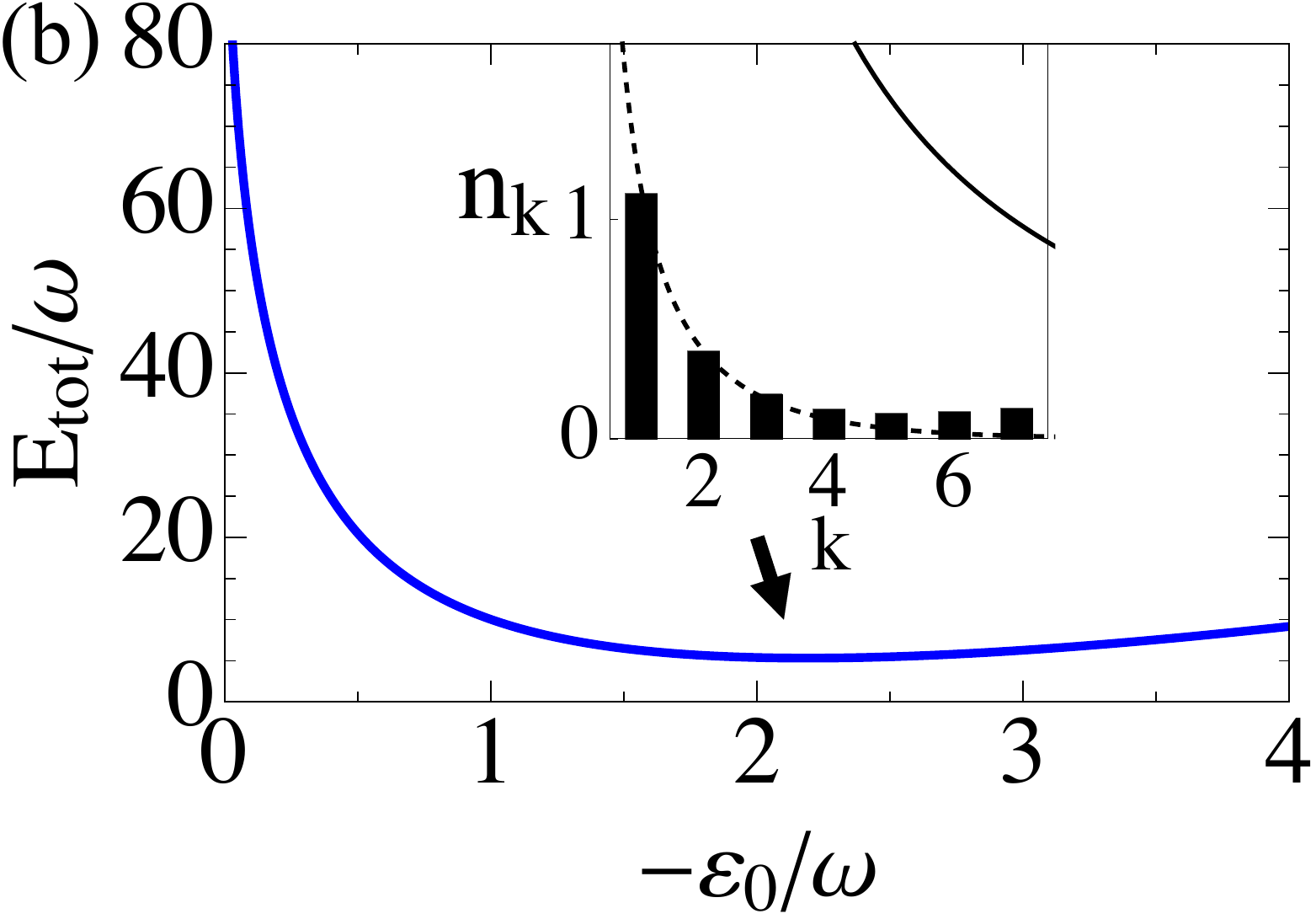} 	
	\caption{The total mechanical energy with $N=7$ modes   
	at $eV=16\omega$ for $\Gamma_s=0.1\omega$ (a) and $\Gamma_s=5\omega$ (b). 
	The insets show the occupation of each modes $k$ at specific points of $\varepsilon_0$ specified by the arrows. 
	The solid black line shows the thermal distribution $n_B(\omega_k)$. 
	The dashed line in the inset (b) is the bose distribution $n^*_B(\omega_k)$ with 
	$T^{*} = 1.6 \omega$. Parameters: $\Gamma_n = 0.1\omega$, $T= 10 \omega $.} 
	\label{fig:multimode}
\end{figure}

We assume a low-frequency spectrum  $\omega_k =  k \, \omega$  (i.e.~under sufficiently high-tension). 
We limit the calculations by considering a finite number of modes as we have a natural cut-off given by the temperature: 
high-frequency modes with $\omega_k \agt T$ are close to the ground state.
%
%
As an example, in Fig.~\ref{fig:multimode}, we show the result for the total mechanical energy defined as 
$E_{\text{tot}} = \sum_{k=1}^{7} \omega_k n_{k}$ for  the case for $T=10  \omega$  and  $N=7$ modes.
The non-equilibrium value $n_{k}$ for each modes is calculated by Eq.~\eqref{eq:main_result} for  
$\gamma_0 \ll \gamma_{\textrm{NR}}$.

In Fig.~\ref{fig:multimode}(a) we consider the regime of weak coupling between the dot and the superconductor
$(\Gamma_s  \ll \omega_k)$, namely the regime of cooling by resonance. 
In this case, by matching the resonance condition $2 |\varepsilon_0| = \omega_k$, 
one can obtain cooling of each individual mode, as for instance for $k=1$ or $k=6$, 
whereas  the rest of the modes  are approximately at the thermal equilibrium.

In Fig.~\ref{fig:multimode}(b) we consider the regime of strong coupling between the dot and the superconductor
 $(\Gamma_s \gg \omega_k)$, namely the regime of cooling by interference.  
In this case, several modes of the resonator can be cooled close to the ground state simultaneously.
Notice that the non-equilibrium distribution of the modes does not correspond to $n^{*}_B(\omega_k)$ with a common 
effective temperature  $T^{*}$ (e.g. see the tail of the fitting curve in the inset of Fig.~\ref{fig:multimode}(b)).     
%
Indeed the phonon occupation for each mode $n_k$ is an interpolation (with frequency dependent coefficients) 
between $n_B(\omega_k)$ resulting from NRs  and the algebraic function $n_{\textrm{AR}}\sim n_{\textrm{opt}}={\Gamma_n^2/\omega_k^2}$ 
resulting from ARs.

%
%
%

\begin{figure}[tbp]
	\includegraphics[scale=0.37,angle=0.]{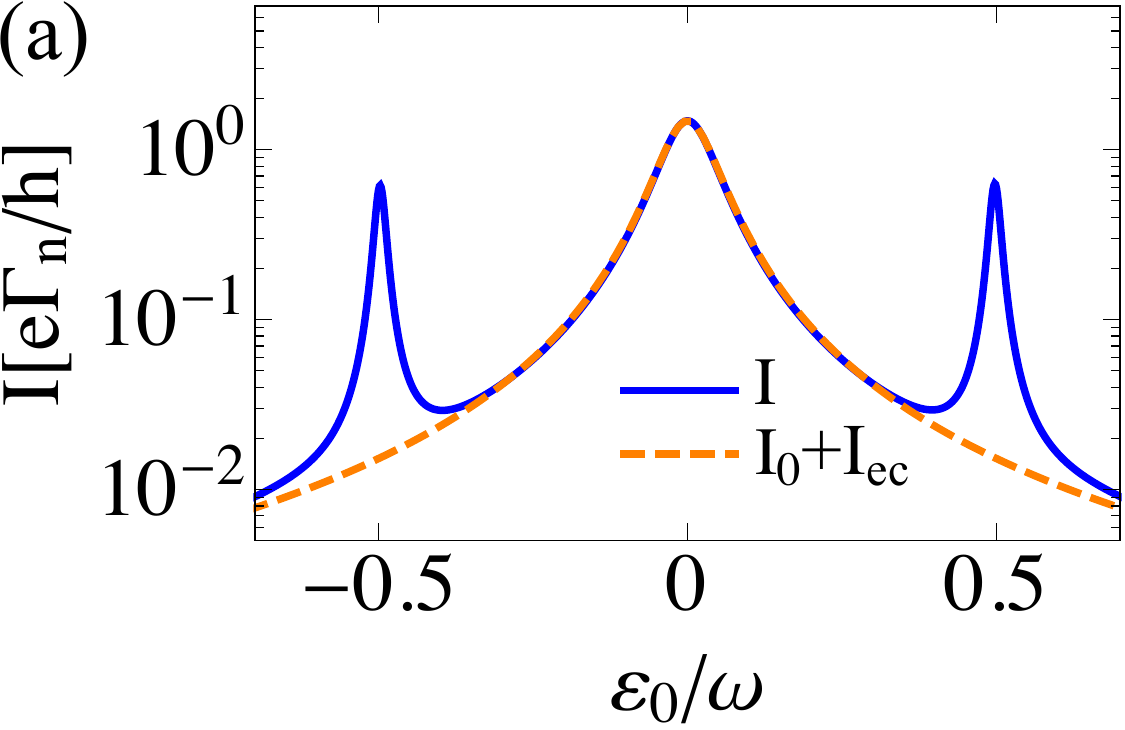}
	\includegraphics[scale=0.37,angle=0.]{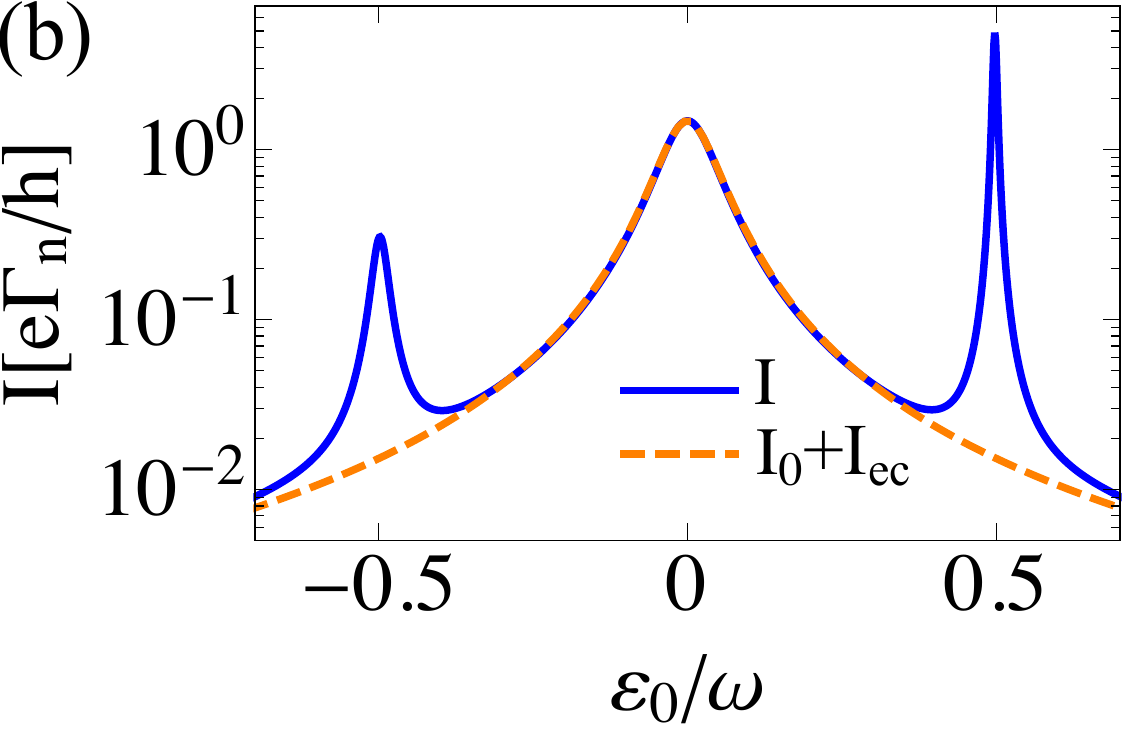} 
%
	\caption{ Total current $I$ (solid blue line) and elastic current (dashed orange line) 
	as function of $\varepsilon_0$ .
	In (a) the case of the thermal equilibrated oscillator $\bar{n}=n_B$.
	In (b) the oscillator is in a non-equilibrium state with $\lambda=0.02\omega$ and  $\gamma_0/\omega=10^{-4}$.
	The ratio between the areas underlying the two peaks is related to the phonon occupation.
	Other parameters: $\Gamma_n=0.01\omega$, $\Gamma_s=0.04\omega$, $T= 10 \omega$, $eV=5\omega$.}    
	\label{fig:current}
\end{figure}

%
%
%
%
%
%
%
%
%
{\sl DC current.}-
We discuss as example the result for a single mode.
To lowest order in the charge-vibration coupling $\lambda$, the current can be expressed as
$I = I_0+I_{\text{ec}}(\lambda^2) + I_{\text{in}} (\lambda^2) $,
with the elastic current $I_0$, the elastic correction $I_{\text{ec}}$ and the inelastic current $I_{\text{in}}$.
Fig.~\ref{fig:current} shows the current at $eV=5\omega$ as a function of $\varepsilon_0$. 
Beyond a peak at $\varepsilon_0=0$ associated to $I_0+I_{\text{ec}}$, 
two vibrational peaks appear at $\varepsilon_0 = \pm \omega/2$
associated to inelastic ARs with emission or absorption of one phonon.
Similar vibrational sidebands have been observed for molecular vibrational modes but 
under the condition $T < \omega$ (for instance, in suspended CNTQDs,  see Refs.~ \cite{Leturcq:2009br,LeRoy:2004bt,Sapmaz:2006kv})
and in other non suspended devices due to other bosonic modes of the environment \cite{Gramich:2015dk}.
In our case, these peaks are visible in the subgap transport even for 
the temperature of the leads $T \gg \omega$.
Analytic expressions for the $I_0$  and $I_{\text{ec}}$ are given in~\cite{SupMat}, 
here we focus on the inelastic term $I_{\text{in}}$.

%
%
%

As shown in Fig.~\ref{fig:current}, around the region $\varepsilon_0 = \pm \omega/2$ the main contribution to the current is given by the inelastic component, which reads
%
%
%
%
%
%
%
%
%

\begin{equation}
\label{eq:current}
	 I_{\text{in}} / e  = 
	 \left[
	 \bar{n}   \gamma^{+}_{eh} 
	 +
	 (\bar{n}+1)  \gamma^{-}_{eh}   
	 \right] 
	 -
	 \left[
	 \bar{n}     \gamma^{+}_{he}   
	 +
	 (\bar{n}+1) \gamma^{-}_{he}    
	 \right] 
	   \, .
\end{equation}
%
%
%
%
For instance, for positive $eV>0$ and high-voltage limit, the first term of the r.h.s of Eq.~(\ref{eq:current}) is the leading one and it is 
associated to the flux of the incoming electrons, as discussed previously. 
For $\gamma_0 \gg \gamma$, $\bar{n} \simeq n_B(\omega)$ and the peaks are approximately symmetric, Fig.~\ref{fig:current}(a).
In the opposite case $\gamma_0 \ll  \gamma$, 
the oscillator is in the non-equilibrium state  as given by Eq.~(\ref{eq:main_result}) 
and the two peaks are strongly  asymmetric, Fig.~\ref{fig:current}(b).
%
%
Integrating the peaks separately over $\varepsilon_0$ (see Ref.~\cite{SupMat}), we can extract information about the phonon occupation by the ratio 
$
\rho = \Delta I_{l} / \Delta I_{r} \approx  (2\bar{n}_l+1)/(2\bar{n}_r+1)
$
where $\Delta I_{{l,r}}$ are the approximated integrals of the left and right peaks in Fig.~\ref{fig:current} and
 $\bar{n}_{l}$  and $\bar{n}_{r}$ are the phonon occupations around such peaks.

One can verify the non-equilibrium state of the resonator in different ways. 
For example, one can vary the voltage and calculate $\rho (eV)$ for each point: 
at low voltage the resonator is close to the thermal state with $\rho \simeq 1$ whereas at high voltage one expects $\rho \ll 1$.
 Alternatively, one can tune the coupling  $\Gamma_s$ with the superconductor, as in the experimental set up of Ref.~\cite{Benyamini:2014eb}. 

%
Finally,  when many mechanical modes are considered, several peaks appear in the inelastic current with broadening $\Gamma_n$ 
\cite{no-broadening_Gamma_S} and one can repeat the same procedure for determining $\rho(\omega_k)$ associated to each mode $k$.

{\sl Conclusions.}-
We discussed the ground state cooling due to inelastic ARs for a mechanical resonator coupled to a quantum dot. 
%
We showed that the destructive interference in the ARs with phonon emission allows for cooling of several mechanical modes. 
%
%
Our proposal is well within the reach of the  state of art for carbon-based NEMS.
The set-up with hybrid contacts in Fig.~\ref{fig:schema_system_idea}(a)
can be experimentally implemented \cite{Schindele:2014fm,Gramich:2015dk} as well as 
%
the strong electromechanical coupling regime for flexural modes  \cite{Benyamini:2014eb,Steele:2009ko,Lassagne:2009fg,Eichler:2012ju}  
such that the intrinsic damping $\gamma_0$  can be much smaller than the electromechanical one  $\gamma \gg \gamma_0$.

\acknowledgments 
We acknowledge A. Armour for interesting discussions and for a critical reading of the manuscript. 
We also thank S. Girvin, A.K. Huettel and A. Bachtold for useful comments. 
This research was supported by the Zukunftskolleg of the University of Konstanz 
and by the DFG through SFB 767.

\bibliography{references}

\begin{thebibliography}{47}
\bibitem [{no-()}]{no-broadening_Gamma_S}%
  \BibitemOpen
  \href@noop {} {}\bibinfo {note} {Note that the coupling $\Gamma_s$ to
  theerconductor does not produce any broadening.}\BibitemShut {Stop}%
\end{thebibliography}

\end{document}